﻿

\documentclass[twocolumn]{aastex62}
\usepackage{amsmath}

\shorttitle{Green Valley Galaxies in Different Environments}
\shortauthors{Jian et al.}

\begin{document}

\title{Redshift Evolution of Green Valley Galaxies in Different Environments from the Hyper Suprime-Cam Survey}

\correspondingauthor{Hung-Yu Jian}
\email{hyjian@asiaa.sinica.edu.tw}

\author{Hung-Yu Jian}
\affiliation{Institute of Astronomy \& Astrophysics, Academia Sinica, 106, Taipei, Taiwan}

\author{Lihwai Lin}
\affiliation{Institute of Astronomy \& Astrophysics, Academia Sinica, 106, Taipei, Taiwan}

\author{Yusei Koyama}
\affiliation{Subaru Telescope, National Astronomical Observatory of Japan, 650 North A’ohoku Place, Hilo, HI 96720, USA}
\affiliation{Department of Astronomical Science, SOKENDAI, Osawa, Mitaka, Tokyo 181-8588, Japan}

\author{Ichi Tanaka}
\affiliation{Subaru Telescope, National Astronomical Observatory of Japan, 650 North A’ohoku Place, Hilo, HI 96720, USA}

\author{Keiichi Umetsu}
\affiliation{Institute of Astronomy \& Astrophysics, Academia Sinica, 106, Taipei, Taiwan}

\author{Bau-Ching Hsieh}
\affiliation{Institute of Astronomy \& Astrophysics, Academia Sinica, 106, Taipei, Taiwan}

\author{Yuichi Higuchi}
\affiliation{Institute of Astronomy \& Astrophysics, Academia Sinica, 106, Taipei, Taiwan}

\author{Masamune Oguri}
\affiliation{Research Center for the Early Universe, University of Tokyo, Tokyo 113-0033, Japan}
\affiliation{Department of Physics, University of Tokyo, Tokyo 113-0033, Japan}
\affiliation{Kavli Institute for the Physics and Mathematics of the Universe (Kavli IPMU, WPI), Universityof Tokyo, Chiba 277-8582, Japan}

\author{Surhud More}
\affiliation{Kavli Institute for the Physics and Mathematics of the Universe (Kavli IPMU, WPI), Universityof Tokyo, Chiba 277-8582, Japan}

\author{Yutaka Komiyama}
\affiliation{National Astronomical Observatory of Japan,
2-21-1 Osawa, Mitaka, Tokyo 181-8588, Japan}
\affiliation{Graduate University for Advanced Studies (SOKENDAI),
2-21-1 Osawa, Mitaka, Tokyo 181-8588, Japan}

\author{Tadayuki Kodama}
\affiliation{National Astronomical Observatory of Japan, Osawa 2-21-1, Mitaka, Tokyo 181-8588, Japan}

\author{Atsushi J. Nishizawa}
\affiliation{Institutie for Advanced Research, Nagoya University, Furocho, Nagoya 464-8602, Japan}

\author{Yu-Yen Chang}
\affiliation{Institute of Astronomy \& Astrophysics, Academia Sinica, 106, Taipei, Taiwan}


\begin{abstract}
Green valley galaxies represent the population that is likely to transition from the star-forming to the quiescent phases. To investigate the role of the environment in quenching star formation, we use the wide-field data from the Hyper Suprime-Cam Strategic Subaru Proposal survey to quantify the frequency of green valley galaxies in different environments and their redshift evolution. We find that the green valley fraction, in general, is less than 20\% in any redshift and environment. The green valley fraction, when normalized to the total population, is higher in the field than that in groups or clusters and decreases with a decreasing redshift and increasing mass. The lower fraction of transitional galaxies in denser environments could be a consequence of the lack of star-forming galaxies, which could be the progenitors of green valley galaxies. To assess the effect of the environment on star formation quenching, we define the effective green valley fraction as the ratio of the number of green valley galaxies to that of nonquiescent galaxies only. The effective green valley fraction for field galaxies is lower than that for group or cluster galaxies, which reveals a strong positive mass dependence and mild redshift evolution. Moreover, the specific star formation rate (sSFR) is reduced by 0.1\textendash 0.3 dex in groups or clusters. Our results thus imply that an ongoing slow quenching process has been acting in the dense environment since $z$ $\sim$ 1.

\end{abstract}



\section{Introduction} \label{sec:intro}

Galaxies reveal a bimodal distribution and can be classified into two distinct populations. The first class comprises passively evolving red galaxies with old stellar populations and early-type morphologies, preferentially residing in high-density regions (e.g., groups and clusters). The other class comprises blue star-forming galaxies with late-type morphologies, primarily found in low-density environments. The galaxy color or morphology bimodality has been observed in the local universe \citep{str01,bla03,bal04} and at a high redshift $z$ $\sim$ 1 or above \citep{bel04,wei05,wyd07,whi11,lin12}.

The relatively sparse region located between these two populations in the color\textendash magnitude or star formation rate (SFR)\textendash stellar mass ($M_{*}$) diagrams has been argued as the overlapping tail of two modeled Gaussian distributions for the two corresponding populations \citep{bal04,tay15}. However, this zone is also considered the crossroad of galaxy evolution. A third class of galaxies called the ``green valley'' population \citep{mar07,wyd07} exists, which has received research interest in recent years. The green valley galaxies are believed to be likely sources for the transition from the active star-forming to quiescent phases [see a review by \cite{sal14}]. Under the above scenario, investigating the properties and abundance of these green valley galaxies in different environments is expected to elucidate the star formation quenching mechanism.

Many studies have attempted to determine the possible dominant quenching mechanisms by measuring the quenching timescales for galaxies, characterizing the spatial sequence of quenching, or mapping the molecular gas contents of galaxies under quenching \citep[e.g.,][]{mcc08,lot10,wet12,wet13,hai13,lin14,muz14,sch14,tar14,tra16,cro17,jian17,del18,row18,lin19}. For example, \cite{lin17} studied the cold molecular gas content of green valley galaxies by using ALMA and found that green valley galaxies, which are believed to be in the transition phase, have lower star formation efficiency than the normal star-forming galaxies, similar to the case of poststarburst galaxies \citep{sue17,fre18}. The results of these studies imply that the quenching may not necessarily involve complete removal of the cold gas of galaxies. More recently, \cite{lin19} characterized the spatial sequence of quenching for approximately  3000 galaxies selected from the SDSS-IV Mapping Nearby Galaxies at Apache Point Observatory (MaNGA;  \citealt*{bun15}) and concluded that the inside-out quenching is the dominant quenching mode in different environments, even in massive halos.

According to the quenching timescales, the physical processes can be divided into two broad categories: slow and fast mechanisms. The slow processes quench the star formation of galaxies over a timescale of 1 Gyr or more, such as morphological quenching \citep{mar09} and strangulation \citep{lar80,bal00}. By contrast, the fast quenching process can halt the star formation in galaxies within a relatively short period of less than 1 Gyr [e.g., mergers \citep{mih94} and ram pressure stripping \citep{gun72}].

In practice, observational results lead to diverse conclusions on the transition timescale. For instance, \cite{sch14} traced the evolution of early- and late-type galaxies through the green valley and found that star formation of early type-galaxies rapidly diminishes in time $<$ 250 Myr accompanied by a notable morphological transformation. By contrast, late-type galaxies undergo gradual quenching in star formation over approximately 1 Gyr without significant morphological change. Based on observations, \citet{sal14}  proposed that the green valley galaxies evolve quasi-statically (i.e., the majority of galaxies now present in the green valley was partially quenched in the past and currently undergoing a passive and slow evolution process). \cite{wet13} found a ``delayed-then-rapid'' quenching scenario in which satellite star formation remains unquenched for 2\textendash 4 Gyr after accretion, and a rapid quenching of star formation starts afterward at $<$0.8 Gyr. Recently, \cite{coe18} used an SDSS galaxy sample to probe the properties of passive, star-forming, and green valley galaxies in four environments, namely the field, groups, outskirts, and core of X-ray clusters. They found that quenching timescales for green valley galaxies in the dense environments are 1.2\textendash 1.8 Gyr, which is generally shorter than that for the field clusters ($\sim$ 2 Gyr). In addition to the observational results, \cite{tra16} presented theoretical constraints on the timescale by using the EAGLE cosmological hydrodynamical simulation. They discovered that the transition time for galaxies to pass the green valley is approximately 2 Gyr, regardless of the physical quenching mechanism.

In our previous work \citep{jian18}, we used an earlier version of the Hyper Suprime-Cam (HSC) Cluster finding algorithm based on multiband identification of red-sequence galaxies (CAMIRA) cluster catalog \citep{ogu18} and the photo-$z$ galaxy catalog \citep{tan18}, both of which were constructed on the basis of HSC S16A wide-field data, to probe the star-forming activity of galaxies in different environments over 0.2 $<$ $z$ $<$ 1.1. We found that star-forming galaxies in groups or clusters exhibit a systematic reduction in the specific star formation rate (sSFR) by 0.1\textendash 0.3 compared with that in the field, and the offsets depend mildly on redshift over the redshift range probed, indicating a universal slow quenching mechanism acting in the dense environments since $z$ $\sim$ 1.1. To further investigate this finding, in the present study, we use the HSC S17A wide field dataset, an internal data release, and explore the distribution of galaxy sSFR and the fraction of green valley galaxies in different surroundings as well as their redshift evolutions to understand the role of the environment at various epochs and to determine the possible dominant quenching mechanisms operating in dense environments.

The remainder of the paper is organized as follows. In Section 2, we briefly describe the study data and the sample selection and analysis methods. In Section 3, we present the main results and discuss the distribution of sSFR and the redshift and mass dependence of the green valley galaxy fraction in the field, group, and cluster environments. Finally, we present conclusions in Section 4. Throughout this paper, we adopt the following cosmology: \textit{H}$_0$ = 100$h$~km s$^{-1}$ Mpc$^{-1}$, $\Omega_{\rm m} = 0.3$, and $\Omega_{\Lambda } = 0.7$. We adopt the Hubble constant $h$ = 0.7 in the calculation of rest-frame magnitudes. All magnitudes are in the AB system.

\section{Data, Sample Selection, and Method}

\subsection{HSC Galaxy Sample}
The HSC Survey is a 300-night Strategic Survey Program that uses 1.77 square degrees of Hyper Suprime-Cam to collect broadband images in \textit{grizy} bands and to detect emission-line objects at high redshifts through four narrowband filters \citep{aih18,fur18,kaw18,kom18,miy18}. Revealing the nature of dark matter and dark energy as well as studying the evolution of galaxies are two main research goals. Three-layered imaging, namely wide, deep, and ultradeep, is conducted. The ``wide'' layer is expected to reach a depth of $r \sim$ 26 mag over the target coverage of 1400 deg$^2$. Moreover, the ``deep'' layer consists of four separate fields with a target depth of $r \sim$ 27 mag over 27 deg$^2$, whereas the ``ultradeep'' layer covers two areas, with a target depth of $r \sim$ 28 mag in 3.5 deg$^2$. The first HSC public data was released in 2017 and presented in \cite{aih18}. 

In this study, our dataset is based on data products internally released as S17A in 2017 September, containing imaging data observed from 2014 March to 2017 May. The full-color full-depth area in the wide survey covers $\sim$ 225 deg$^2$. This release data are processed using hscPipe \citep{bos18} (version 5.4), which is based on the Large Synoptic Survey Telescope pipeline \citep{ive08,axe10,jur15}. In addition, from the S17A release note, the quality assurance test results reveal that the astrometry is effective at 10\textendash 20 mas against GAIA; however, there are small-scale ($\sim$ 1 deg) regions with larger systematic offsets. The photometry is precise down to $\sim$ 0.01\textendash 0.02 mag; however, a pixel-to-pixel spatial variation of zero-points at a level of $\sim$2\% exists.

\subsection{CAMIRA Groups/Clusters}
CAMIRA, developed by \cite{ogu14}, adopts the stellar population synthesis model of \cite{bru03} and fits all photometric galaxies for an arbitrary set of bandpass filters. It also computes the likelihood of being red-sequence galaxies as a function of redshift. The detailed methodology of the CAMIRA algorithm can be found in \cite{ogu14}. Unlike our previous study \citep{jian18}, which used the HSC Wide S16A CAMIRA catalog released in 2017 \citep{ogu18} for the analysis, in this study, we use the new optically selected HSC Wide S17A CAMIRA wide catalog in full colors and full depth, based on the S17A photometry from S17A internal release. The catalog contains 7294 clusters/groups with richness $N_{\textrm{mem}}$ $>$ 10 in the redshift range of 0.1 $<$ $z$ $<$ 1.38. The catalog is further divided into two richness levels (i.e., 10 $<$ $N_{\textrm{mem}}$$<$ 25 and $N_{\textrm{mem}}$ $>$ 25) and three redshift bins, namely 0.2 $< z <$ 0.5, 0.5 $< z <$ 0.8, and 0.8 $< z <$ 1.1. Based on Equation 40 in \citet{ogu14}, $N_{\textrm{mem}}$ = 10 and 25 correspond to the virial halo mass log$_{10}$($M_{\textrm{vir}}/h^{-1}$ $M_{\odot}$) $\sim$ 13.61 $\pm$ 0.13 and 14.19 $\pm$ 0.02, respectively. The total number of adopted groups or clusters in this analysis is 7292 in the probed redshift of 0.2 $< z < $1.1. Table~\ref{tab1} lists the numbers of groups and clusters in three corresponding redshift ranges. The number increase for clusters is approximately 60\%, whereas for groups, it is doubled compared with the S16A CAMIRA catalog.   

\begin{deluxetable}{cccc}
\tablenum{1}
\tablecaption{S17A CAMIRA Cluster Catalog \label{tab1}}
\tablehead{
\colhead{Redshift} & \colhead{$z_{\textrm{mean}}$} & \colhead{Group} & \colhead{Cluster} \\   
\colhead{} &  & \colhead{$10 < N_{\textrm{mem}}< 25$} & \colhead{$N_{\textrm{mem}}> 25$} \\
\colhead{} &  & \colhead{$M_{\textrm{vir}}$/$h ^{-1}$ $M_{\odot}$ = $10^{13.6-14.2}$} &  \colhead{$10^{> 14.2}$} 
}

\startdata
 0.2 $< z <$ 0.5  & 0.35 & 1667 & 235  \\
 0.5 $< z <$ 0.8  & 0.67 & 2503 & 202  \\
 0.8 $< z <$ 1.1  & 0.94 & 2558 & 127  \\
\enddata

 
\end{deluxetable}

\subsection{Photo-$z$ catalog and stellar mass estimation}

The photometric redshift and stellar mass are estimated using the direct empirical photometric method \citep[DEmP;][]{hsi14}. DEmP is an empirical-fitting code and designed to minimize the effects of two main problems in conventional empirical-fitting methods (i.e., the choice of the proper form for the fit function and the bias in the best-fit coefficients caused by objects with high population density).  DEmP adopts a method called ``regional polynomial fitting'' to dynamically select a local subset, which consists of training galaxies with magnitudes and colors that are closest to the input galaxy, to derive the relation represented by a first-order polynomial function between redshift and photometry for that input galaxy. In this manner, an arbitrary variation in the color\textendash magnitude space can be represented by multiple line segments without the need to choose a complicated fit function that satisfies all data. In addition, DEmP uses a uniformly weighted training set by artificially making the number of galaxies in each redshift bin in the training set the same to alleviate the bias in the best-fit coefficients caused by objects with high population density in the relevant parameter space. The redshift and stellar mass estimation using DEmP are described in detail in \cite{hsi14}.

 The photometric redshift and stellar mass are computed independently using DEmP. The training set for the photo-$z$ computation consists of several redshift catalogs, including  approximately $1.7\times10^{5}$ sources from the COSMOS2015 multiband photo-$z$ catalog \citep{lai16}, assuming a \cite{cha03} initial mass function, and  approximately $3.7\times10^{4}$ sources from public spectroscopic samples based on SDSS DR14, DEEP3 DR4,  PRIMUS DR1, VIPERS PDR1, VVDS, GAMA DR2, WiggleZ DR1, zCOSMOS DR3, UDSz, 3D-HST v4.1.5, and FMOS-COSMOS v1.0 \citep[see][and references therein]{tan18}. Notably, all the objects in the training set contain redshift information, but only objects from the COSMOS2015 catalog have stellar-mass measurements. The redshift (and stellar mass) for each object is calculated using the 40 nearest neighbors in the nine-dimensional space (five magnitude axes and $g-r$, $r-i$, $i-z$, and $z-y$ color axes) with a linear function. 
 
 The DEmP photo-$z$ catalog provides the photometric redshift and stellar mass for galaxies even if their photometry is available in one band only. For the sample with faint objects (i.e., including $i > 25$ galaxies), the photo-$z$ dispersion $\sigma$ defined as
\begin{equation}\label{eq1}
\sigma = 1.48 \times \textrm{MAD}(\Delta z),
\end{equation}
where $\Delta z$ = $z_{\textrm{phot}}$ - $z_{\textrm{spec}}$ and MAD is the median absolute deviation, is approximately 0.052 and the outlier rate is 28$\%$ when the best point estimator is used, which is defined as an optimal estimate to minimize a loss function (the detailed description can be found in the photo-$z$ release paper by \citealt*{tan18}). However, the stellar mass corresponding to COSMOS masses shows a mean ($\Delta$log$M_*$ = log$M^{\textrm{hsc}}_*$ - log$M^{\textrm{cosmos}}_*$) of -0.02 dex and a scatter of 0.2 dex. For the present analysis, we select sources with i\_extendedness\_value = 1, which indicates the condition of the galaxy extendedness in the $i$ band to be the extended sources, and afterward set photoz\_std\_best, which denotes the standard deviation around photo-$z$, as $\leq$0.3 to select galaxies with favorable photo-$z$. The total number of galaxies in the photo-$z$ sample is $\sim$ 19 million for galaxies with i $<$ 26 and photo-$z \leq$ 1.1.

\subsection{Methods}\label{method}

\begin{figure*}
\includegraphics[scale=1.4]{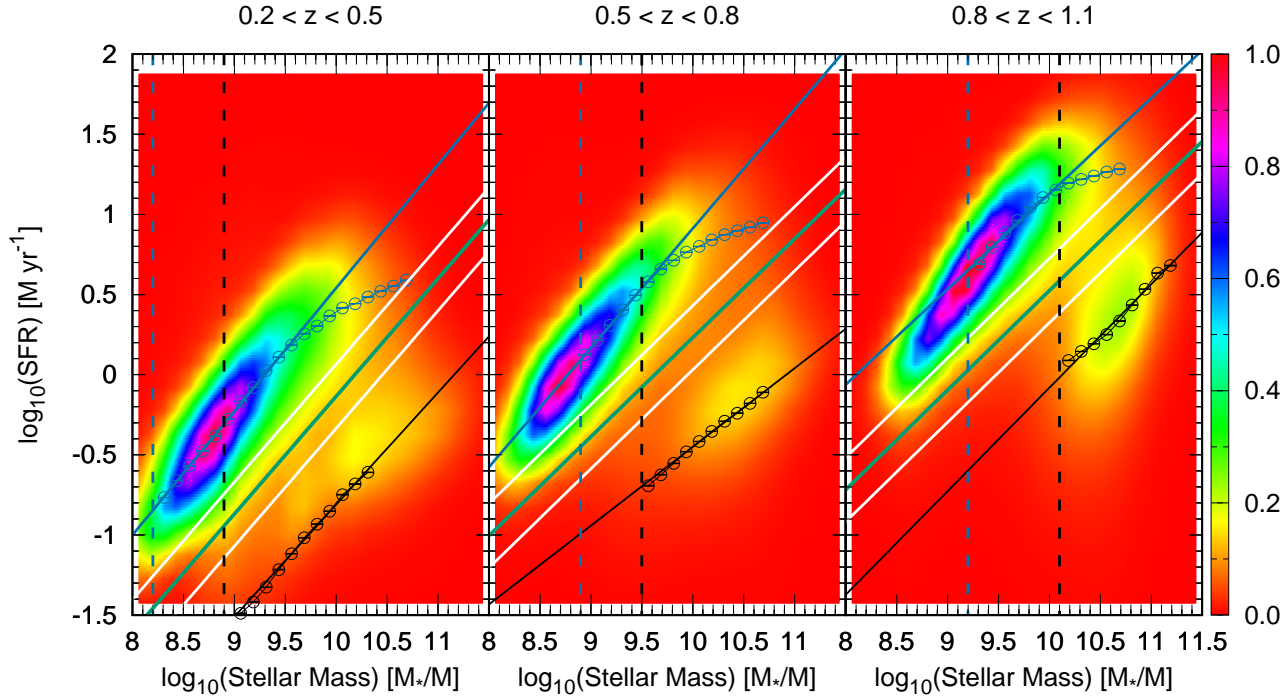}
\caption{Definition of the green valley galaxies by using all galaxies on the SFR$-M_{*}$ plane in three redshift ranges. The blue and black open circles in each panel represent the median sSFR of star-forming and quiescent galaxies, respectively. The blue and black solid lines denote the best-fit results of the median sSFR of the star-forming and quiescent galaxies from different fitting mass ranges described in Section~\ref{gv}, respectively. The green line is the average of the blue and black lines and represents the central ridge of the green valley zone. The region enclosed by two parallel white lines 0.2 dex above and below the green line defines the transition area occupied by the green valley galaxies. Moreover, the vertical blue and black dashed lines denote the mass completeness limits for star-forming and quiescent galaxies in the redshift range, respectively.}
\label{f1}
\end{figure*}

To extract properties of cluster galaxies, we follow the procedure described in \citet{jian18}. For each cluster, galaxies around the cluster center with redshift 1 $\sigma$ above or below the cluster redshift are projected onto a 2D plane, where $\sigma$ is the photo-$z$ dispersion of the galaxy sample defined in Equation~\ref{eq1}, and is approximately 0.052 in this work. For these projected galaxies, we replace individual photo-$z$s with the same cluster redshift when computing the rest-frame $B$ magnitude $M_{B}$ and $(U-B)_0$ color for each galaxy by using the K-correction based on empirical templates from \cite{kin96}. A detailed description is provided in \cite{jian18} and the references therein. We then employ the formula in \cite{mos12}, i.e.
\begin{equation}\label{eq2}
\begin{split}
 \textrm{SFR}[M_{\odot} \ yr^{-1}] = \\ & 0.318-0.424 \textrm{M}_{B} \\
& +2.925 (\textrm{U-B})_0-2.603 {(\textrm{U-B})_0}^{2},
\end{split}
\end{equation}
where $M_{B}$ is the rest-frame $B$ magnitude and $(U-B)_0$ is the rest-frame $U-B$ color. The SFR estimation yields an uncertainty of approximately 0.19 dex for star-forming and 0.47 dex for quiescent galaxies, with a mean residual offset of $-$0.02 dex. We note that the SFR calibration is only applicable in deviation from the global SFR behavior of a large sample and is not suitable for use on an individual galaxy basis \citep{mos12}. 

Because CAMIRA is a cluster finder based only on red-sequence galaxies, it does not recover all the galaxy members of groups/clusters. Therefore, we perform galaxy stacking combined with the background/foreground subtraction technique to investigate the statistical properties of cluster galaxies. Galaxies within a comoving projection radius $r_p$ of 1.5 \textrm{Mpc} from the cluster center, referred to as the contaminated cluster sample, constitute both cluster members and foreground/background galaxies, whereas galaxies in an annulus between an $r_p$ of 8.0 and 10.0 \textrm{Mpc} form the field sample. For a given parameter space (e.g., SFR vs. $M_{*}$), we compute the number counts for the contaminated cluster and field samples, separately, by normalizing the area. We can then recover the statistical properties of cluster galaxies by subtracting the number count of the field sample from the corresponding number count of the contaminated cluster sample. 

Our method for the background subtraction is similar to those of \cite{pim02} and \cite{val12}, except that we do not apply a correction for grids with a negative number of galaxies. The negative grids originate from both the nonzero and zero sources in the contaminated sample. The ratio of the number from negative cells with zero sources to that from total cells is roughly 12\%, and for the nonzero sources, it is approximately 2\%. In either case, we keep the negative values and take them into account when we compute galaxy properties, such as the fraction and the median sSFR, at a given stellar mass bin.

For a fair comparison, the radius should be normalized to the cluster radius to reduce its mass and redshift dependence. However, the size of radius is not well defined in the cluster catalog. For simplicity, we thus set the radius as 1.5 Mpc at all redshift and cluster mass. Additionally, for the estimation of the mass completeness limits, we follow the approach by \cite{ilb10}, which defines the low stellar mass limits as the masses with the fraction of galaxies fainter than $i >$ 24.0 at a limit of 30$\%$. We find that the mass limits are log$_{10}$($M_*$ /$M_\odot$) = 8.2 (8.9), 8.9 (9.5), and 9.2 (10.1) for star-forming (quiescent) galaxies in a redshift range of 0.2\textendash 0.5, 0.5\textendash 0.8, and 0.8\textendash 1.1, respectively.

\begin{deluxetable*}{ccccccccc}
\tablenum{2}
\tablecaption{Best-fit Parameters for Star-forming Main Sequence, Red Sequence, and Green Valley\label{tab2}}
\tablehead{
\colhead{Redshift} & \multicolumn{2}{c}{Star-forming Main Sequence} & \colhead{} & \multicolumn{2}{c}{Red Sequence}  &  \colhead{} & \multicolumn{2}{c}{Green Valley} \\ 
\cline{2-3}\cline{5-6}\cline{8-9}
\colhead{} & \colhead{$\alpha$\tablenotemark{a}} & \colhead{$\beta$\tablenotemark{a}} &  \colhead{} & \colhead{$\alpha$\tablenotemark{a}} & \colhead{$\beta$\tablenotemark{a}} &  \colhead{} & \colhead{$\alpha$\tablenotemark{a}} & \colhead{$\beta$\tablenotemark{a}} \\   
\cline{2-3}\cline{5-6}\cline{8-9} 
}

\startdata
 0.2 $< z <$ 0.5   & 0.77 $\pm$ 0.01 & -7.16 $\pm$ 0.08  & & 0.70 $\pm$ 0.01 & -7.81 $\pm$ 0.13  & & 0.73 $\pm$ 0.01 & -7.48 $\pm$ 0.15  \\
 0.5 $< z <$ 0.8   & 0.74 $\pm$ 0.01 & -6.53 $\pm$ 0.04  & & 0.49 $\pm$ 0.01 & -5.36 $\pm$ 0.07  & & 0.62 $\pm$ 0.01 & -5.94 $\pm$ 0.08 \\
 0.8 $< z <$ 1.1   & 0.60 $\pm$ 0.03 & -4.83 $\pm$ 0.33  & & 0.65 $\pm$ 0.03 & -6.54 $\pm$ 0.33  & & 0.62 $\pm$ 0.04 & -5.68 $\pm$ 0.47 \\
\enddata
\tablenotetext{a}{$\alpha$ and $\beta$ are the fitting slope and amplitude for the fitting formula, $log_{10}(\rm{SFR}/M_{\odot}  ~ yr^{-1}) = \alpha ~ log_{10}(M_*/M_{\odot}) + \beta$, respectively.}
\end{deluxetable*}

\subsection{Definition of green valley (or transition) galaxies}\label{gv}

In this study, the SFR$-M_{*}$ diagram is used to define green valley galaxies in a manner similar to the physically motivated definition adopted in \cite{pan17}. \cite{pan17} first determined the star-forming main sequence on the SFR$-M_{*}$ plane and then defined the green valley (or transition) region in a range from 1.5$\sigma$ to 3.5$\sigma$ below the star-forming median line, where $\sigma$ is the standard deviation of the SFR for star-forming galaxies. Consequently, the transition region will evolve with the star-forming main sequence or with time. Unlike defining a fixed transition zone with respect to redshift, the evolving green valley zone avoids selecting star-forming low-$z$ galaxies or quiescent high-$z$ galaxies as the green valley galaxies \citep{pan17}.

Similarly, we use the SFR$-M_{*}$ diagram to define the green valley zone. We stack all galaxies from the sample on the SFR$-M_{*}$ plane to determine the star-forming main sequence and quiescent population. The two populations are then fitted with a linear function $log_{10}(\rm{SFR}/M_{\odot}  ~ yr^{-1}) = \alpha ~ log_{10}(M_*/M_{\odot}) + \beta$ for three redshift bins, separately. The mass ranges used for the line fitting are 8.3\textendash 9.5 (9.2\textendash 10.3), 8.5\textendash 9.5 (9.7\textendash 10.8), and 9.2\textendash 10.2 (10.1\textendash 11.2) in a unit of log$(M_*/M_\odot)$ at low, medium, and high redshifts, respectively, for the star-forming main sequence (quiescent population).

\begin{figure*}
\includegraphics[scale=1.3]{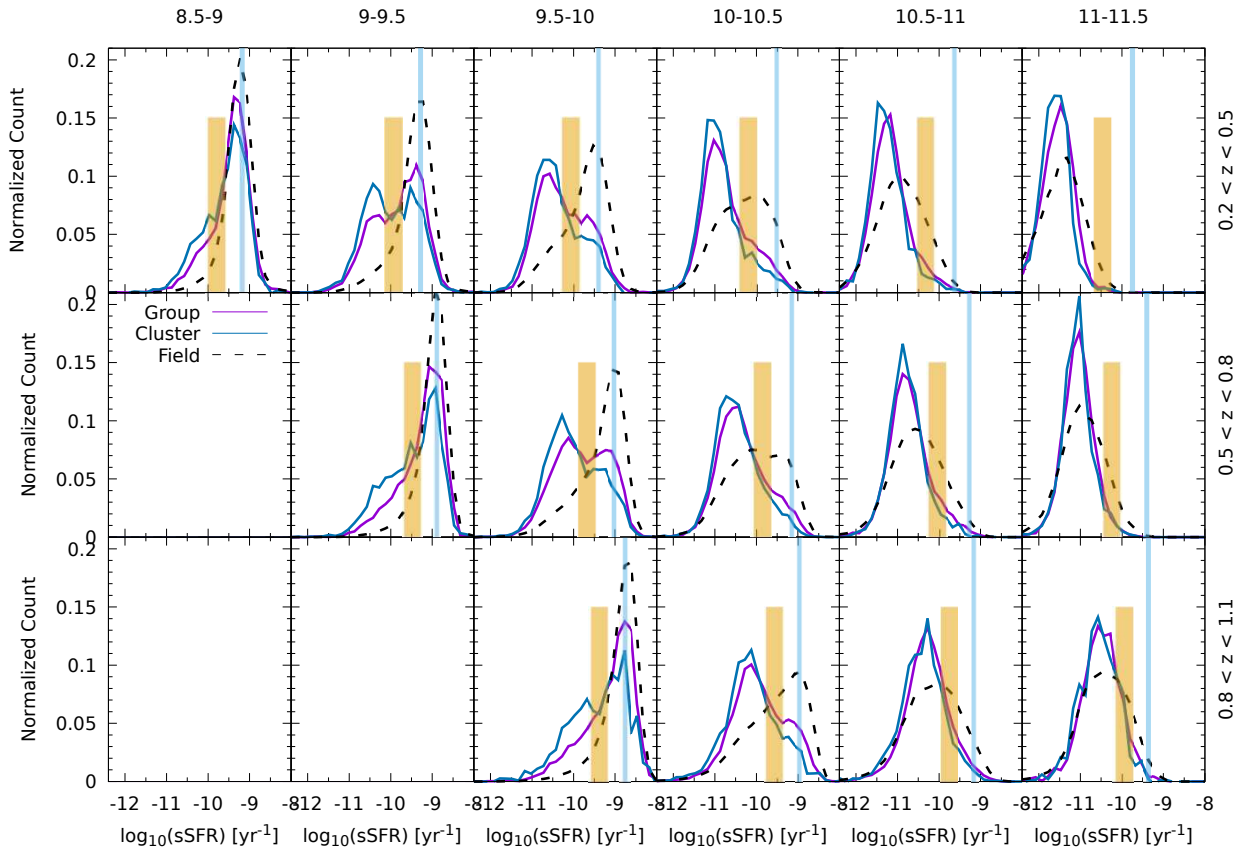}
\caption{Normalized distribution of galaxy sSFR in different environments as functions of stellar mass and redshift. In each panel, the solid purple and blue lines mark the sSFR distribution of group and cluster galaxies, respectively, whereas the black dashed line represents field galaxies. Moreover, the light blue vertical line denotes the median sSFR of star-forming field galaxies, and the gold shaded area defines the green valley region. Three subpanels are empty due to the mass incompleteness at that redshift bin. The red galaxies in the leftmost panel (or the least massive bin) in the redshift range from 0.8 to 1.1 are below the mass completeness limit. The numbers at the top of the subpanel columns in the figure indicate the lower and upper limit masses in a unit of log$_{10}(M_*/M_{\odot})$. From the plot, it is evident that our definition of the green valley region can be applied to the valley of the distribution.}
\label{f2}
\end{figure*}

\begin{deluxetable*}{cccccccccc}
\tablenum{3}
\tablecaption{Fraction of galaxies with log$_{10}$ sSFR $<$ $-$10.0} \label{tab3}
\tablehead{
\colhead{Redshift} & \colhead{} & \colhead{Environments} & \colhead{} & \multicolumn6c{Mass Range [$log_{10}(M_*/M_{\odot})$]} \\
\cline{1-1} \cline{3-3} \cline{5-10}
\colhead{} & \colhead{} & \colhead{} & \colhead{} & \colhead{8.5-9.0} &\colhead{$9.0-9.5$} & \colhead{9.5-10.0}  & \colhead{10.0-10.5} &  \colhead{10.5-11.0} &  \colhead{11.0-11.5} 
}

\startdata
                           & & Field       & &  0.042 &  0.114  &  0.310 &  0.649 & 0.941 & 0.999 \\
 0.2 $< z <$ 0.5  & & Groups   & &  0.147 &  0.375  &  0.653 &  0.865 & 0.978 & 1.000 \\
                           & & Clusters  & &  0.227 &  0.498  &  0.753 &  0.914 & 0.988 & 1.000 \\      
 \cline{1-10}
                           & &  Field      & &    &  0.025 &  0.146 &  0.483 & 0.794 & 0.963 \\
 0.5 $< z <$ 0.8  & & Groups   & &    &  0.110 &  0.402 &  0.724 & 0.906 & 0.992 \\
                           & &  Clusters & &   &  0.222 &  0.528 &  0.796 & 0.944 & 0.991 \\    
   \cline{1-10}                                                
                           & & Field       & &   &  &  0.033 &  0.255 & 0.570 & 0.763 \\
 0.8 $< z <$ 1.1  & & Groups   & &   &  &  0.116 &  0.475 & 0.719 & 0.867 \\
                           & & Clusters  & &  &   &  0.213 &  0.591 & 0.777 & 0.896 \\
 \enddata

\end{deluxetable*}

\begin{deluxetable*}{ccccccccccccc}
\tablenum{4}
\tablecaption{Properties as in Figure~\ref{f2}} \label{tab4}
\tablehead{
\colhead{Redshift} & \colhead{} & \colhead{Environments} & \colhead{} & \colhead{Mass Range} & \colhead {} &  \colhead{Quiescent} & \colhead{} & \colhead{Star-forming} & \colhead{} &  \colhead{Green} & \colhead{} & \colhead{Total Galaxy Number\tablenotemark{a}}\\
\cline{1-1} \cline{3-3} \cline{5-5} \cline{7-7} \cline{9-9} \cline{11-11} \cline{13-13}
\colhead{} & \colhead{} & \colhead{} & \colhead{} & \colhead{$log_{10}(M_*/M_{\odot})$} & \colhead{} & \colhead{$f_q$\tablenotemark{b}} & \colhead{} & \colhead{$f_s$\tablenotemark{b}} & \colhead{} & \colhead{$f_g$\tablenotemark{b}} & \colhead{} & \colhead{\#} 
}

\startdata
                           & &                & &  8.5$-$9.0   & &  0.219 & &  0.585 & &  0.196 & & 3.032$\times 10^{3}$ \\
                           & &                & &  9.0$-$9.5   & &  0.439 & &  0.380 & &  0.181 & & 4.487$\times 10^{3}$ \\
                           & & Clusters  & &  9.5$-$10.0  & &  0.640 & &  0.206 & &  0.155 & & 5.245$\times 10^{3}$ \\
                           & &                & & 10.0$-$10.5 & &  0.831 & &  0.080 & &  0.089 & & 4.899$\times 10^{3}$ \\
                           & &                & & 10.5$-$11.0 & &  0.938 & &  0.021 & &  0.042 & & 3.165$\times 10^{3}$ \\
                           & &                & & 11.0$-$11.5 & &  0.986 & &  0.001 & &  0.013 & & 1.231$\times 10^{3}$ \\             
  \cline{3-13}
                           & &                & &  8.5$-$9.0    & &  0.142 & &  0.680 & &  0.178 & & 1.448$\times 10^{4}$ \\  
                           & &                & &  9.0$-$9.5    & &  0.324 & &  0.492 & &  0.184 & & 1.843$\times 10^{4}$ \\
 0.2 $< z <$ 0.5  & & Groups   & &  9.5$-$10.0   & &  0.530 & &  0.290 & &  0.180 & & 1.914$\times 10^{4}$ \\
                           & &                & & 10.0$-$10.5  & &  0.748 & &  0.128 & &  0.125 & & 1.633$\times 10^{4}$ \\
                           & &                & & 10.5$-$11.0  & &  0.906 & &  0.031 & &  0.064 & & 1.123$\times 10^{4}$ \\
                           & &                & & 11.0$-$11.5  & &  0.990 & &  0.001 & &  0.009 & & 4.794$\times 10^{3}$ \\             
  \cline{3-13}
                           & &                & &  8.5$-$9.0   & &  0.040 & &  0.846 & &  0.114 & & 3.275$\times 10^{4}$ \\                                                  
                           & &                & &  9.0$-$9.5   & &  0.087 & &  0.788 & &  0.124 & & 2.683$\times 10^{4}$ \\
                           & &  Field      & &  9.5$-$10.0  & &  0.197 & &  0.632 & &  0.171 & & 1.882$\times 10^{4}$ \\ 
                           & &                & & 10.0$-$10.5 & &  0.437 & &  0.343 & &  0.220 & & 1.255$\times 10^{4}$ \\
                           & &                & & 10.5$-$11.0 & &  0.743 & &  0.088 & &  0.169 & & 7.203$\times 10^{3}$ \\
                           & &                & & 11.0$-$11.5 & &  0.949 & &  0.006 & &  0.045 & & 2.220$\times 10^{3}$ \\                                       
 \cline{1-13}
                           & &                & &  9.0$-$9.5   & &  0.336 & &  0.467 & &  0.197 & & 2.429$\times 10^{3}$ \\ 
                           & &                & &  9.5$-$10.0  & &  0.594 & &  0.246 & &  0.160 & & 3.925$\times 10^{3}$ \\
                           & & Clusters  & & 10.0$-$10.5 & &  0.784 & &  0.092 & &  0.124 & & 4.492$\times 10^{3}$ \\
                           & &                & & 10.5$-$11.0 & &  0.896 & &  0.033 & &  0.071 & & 3.165$\times 10^{3}$ \\
                           & &                & & 11.0$-$11.5 & &  0.953 & &  0.008 & &  0.038 & & 1.110$\times 10^{3}$ \\       
  \cline{3-13}
                           & &                & &  9.0$-$9.5   & &  0.199 & &  0.617 & &  0.184 & & 1.688$\times 10^{4}$ \\    
                           & &               & &  9.5$-$10.0 & &  0.468 & &  0.349 & &  0.183 & & 2.025$\times 10^{4}$ \\
 0.5 $< z <$ 0.8  & & Groups  & & 10.0$-$10.5 & &  0.707 & &  0.152 & &  0.142 & & 2.215$\times 10^{4}$ \\
                           & &               & & 10.5$-$11.0 & &  0.842 & &  0.059 & &  0.099 & & 1.691$\times 10^{4}$ \\
                           & &               & & 11.0$-$11.5 & &  0.951 & &  0.008 & &  0.040 & & 6.272$\times 10^{3}$ \\  
  \cline{3-13}
                           & &                & &  9.0$-$9.5   & &  0.058 & &  0.817 & &  0.125 & & 3.073$\times 10^{4}$ \\   
                           & &               & &  9.5$-$10.0 & &  0.178 & &  0.669 & &  0.153 & & 2.057$\times 10^{4}$ \\
                           & &  Field     & & 10.0$-$10.5 & &  0.458 & &  0.346 & &  0.196 & & 1.501$\times 10^{4}$ \\
                           & &               & & 10.5$-$11.0 & &  0.679 & &  0.136 & &  0.185 & & 9.956$\times 10^{3}$ \\
                           & &               & & 11.0$-$11.5 & &  0.841 & &  0.037 & &  0.122 & & 2.436$\times 10^{3}$ \\ 
 \cline{1-13}
                           & &                & &  9.5$-$10.0   & &  0.386 & &  0.422 & &  0.172 & & 1.018$\times 10^{3}$ \\   
                           & & Clusters  & & 10.0$-$10.5 & &  0.727 & &  0.134 & &  0.139 & & 1.963$\times 10^{3}$ \\
                           & &                & & 10.5$-$11.0 & &  0.813 & &  0.059 & &  0.129 & & 1.950$\times 10^{3}$ \\
                           & &                & & 11.0$-$11.5 & &  0.836 & &  0.031 & &  0.133 & & 5.009$\times 10^{2}$ \\                             
  \cline{3-13}
                           & &                & &  9.5$-$10.0   & &  0.223 & &  0.601 & &  0.176 & & 1.032$\times 10^{4}$ \\   
 0.8 $< z <$ 1.1  & & Groups   & & 10.0$-$10.5 & &  0.609 & &  0.223 & &  0.168 & & 1.444$\times 10^{4}$ \\
                           & &                & & 10.5$-$11.0 & &  0.757 & &  0.084 & &  0.158 & & 1.564$\times 10^{4}$ \\
                           & &               & & 11.0$-$11.5 & &  0.810 & &  0.043 & &  0.148 & & 4.555$\times 10^{3}$ \\                              
  \cline{3-13}
                           & &                & &  9.5$-$10.0   & &  0.075 & &  0.810 & &  0.115 & & 1.527$\times 10^{4}$ \\   
                           & & Field      & & 10.0$-$10.5 & &  0.343 & &  0.464 & &  0.193   & & 1.132$\times 10^{4}$ \\
                           & &                & & 10.5$-$11.0 & &  0.605 & &  0.188 & &  0.207  & & 9.013$\times 10^{3}$ \\
                           & &               & & 11.0$-$11.5 & &  0.702 & &  0.101 & &  0.196   & & 1.967$\times 10^{3}$ \\                              
 \enddata
\tablenotetext{a}{``Total galaxy number'' refers to the number of galaxies after background subtraction.}
\tablenotetext{b}{$f_q$: quiescent fraction, $f_s$: star-forming fraction, and $f_g$: green valley galaxy fraction.}
\end{deluxetable*}

Subsequently, we employ the middle point of these two populations as the reference line. We then use this reference line to separate the star-forming and quiescent populations on the SFR$-M_*$ plane. Because the quiescent galaxies are less concentrated than star-forming ones, the determination of the quiescent sequence is sensitive to galaxies in the green valley region. We thus recommend that galaxies above the reference line constitute the star-forming population, whereas galaxies 0.5, 0.4, and 0 dex below the reference line are quiescent galaxies in the redshift range of 0.2\textendash 0.5, 0.5\textendash 0.8, and 0.8\textendash 1.1, respectively. We then redefine the new star-forming and quiescent sequences and obtain a new reference line. We iterate the procedure until the slope and interception of the reference line converge. In this manner, we determine the ultimate reference line as the ``green valley line.'' We next define the green valley region as the area enclosed by 0.2 dex below and above the green valley line for any redshift. That is, the width of the green zone is fixed and is also redshift independent. Because the green valley line is the middle point of the star-forming and red sequences, the slope and intercept of the green valley line are the averaged $\alpha$ and $\beta$ from two sequence lines. The best-fit coefficients for the star-forming main sequence, red sequences, and green valley line are listed in Table~\ref{tab2}.

When we start the iteration process, the initial threshold for separating the star-forming and quiescent populations at different redshifts is set as a constant log$_{10}$(sSFR) = $-$10.0. To elucidate how the initial threshold values influence our results, we repeat the iteration procedures with different initial values as a test. The result reveals that the final green valley lines are not sensitive to the starting thresholds adopted in the log$_{10}$(sSFR) range between $-$9.5 and $-$10.5. In addition, to understand the effect of the green valley width on the green valley galaxy fraction, we also compute the green valley galaxy fraction by using a width of 0.15 and 0.1 dex, respectively. We find that the green valley galaxy fraction for a width of 0.15 dex results in approximately 20$\%$ reduction and for a width of 0.1 dex leads to  approximately 40$\%$ decrease compared with the fraction for a width of 0.2 dex. Although quantitative differences exist in the green valley fraction, the main trends discussed in this paper remain similar.

Moreover, the SFR estimation for quiescent galaxies has a large scatter of approximately 0.47 dex, but its offset is comparatively small \citep[$\sim$ 0.02 dex;][]{mos12}. In other words, the definition of the red sequence line by using the median SFR of quiescent galaxies should not deviate from that of the real red sequence line, and the difference is likely on order similar to that of the mean residual offset. Namely, the green valley can be clearly defined.

Figure~\ref{f1} shows the color-coded galaxy density plot on the SFR$-M_*$ plane in a redshift range of 0.2 $< z <$ 0.5 (left), 0.5 $< z <$ 0.8 (middle), and 0.8 $< z <$ 1.1 (right) above the corresponding mass completeness limits. In each subpanel, the density is normalized by the maximum density in the cells. The vertical blue and black dashed lines represent the mass completeness limits for the star-forming and quiescent galaxies at the redshift of each panel, respectively. The two parallel white lines above and below the green valley line (the green line) in each panel indicate the green valley zone. The oblique blue and black lines denote the best-fit results for the median SFR of the star-forming and quiescent galaxies, respectively. However, the median SFR of the star-forming and quiescent galaxies are marked by the blue and black open circles, respectively. It is seen that our defined green valley region precisely captures the transition zone between star-forming and quiescent galaxy populations.

\begin{figure*}
\includegraphics[scale=1.3]{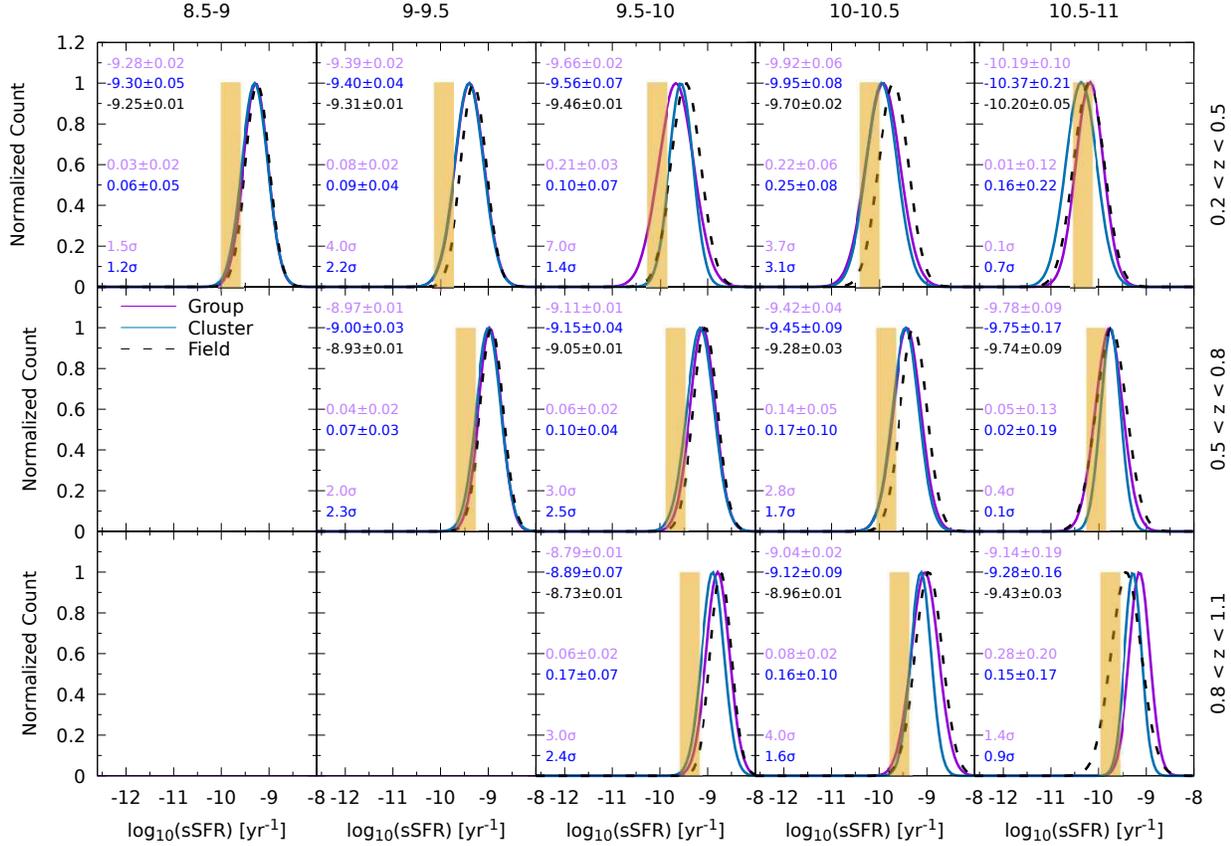}
\caption{Best-fit results of the star-forming population in the group, cluster, and field environments by using two Gaussian profiles. Same as in Figure~\ref{f2}, the numbers at the top of the figure indicate the lower and upper limit masses in a unit of log mass. The numbers in different colors on the top left of each subpanel indicate the peak sSFR value and its asymptotic standard error of star-forming galaxies from the field (black), groups (purple), and clusters (blue), respectively. We also show the sSFR reduction and the significance of the reduction between group and field galaxies (purple) and between cluster and field galaxies (blue) on the middle-left and on the bottom left of each subpanel, respectively. A global sSFR reduction is seen in group or cluster star-forming galaxies in contrast to the field galaxies, consistent with a picture of a slow environmental quenching.}
\label{f3}
\end{figure*}

\begin{deluxetable*}{ccccccccc}
\tablenum{5}
\tablecaption{Results of Two-side K-S test as in Figure~\ref{f3}} \label{tab5}
\tablehead{
\colhead{Redshift} & \colhead{} & \colhead{Environments} & \colhead{} & \colhead{Mass Range} & \colhead {} &  \colhead{K-S Statistic\tablenotemark{a}} & \colhead{} & \colhead{p-value\tablenotemark{a}} \\
\cline{1-1} \cline{3-3} \cline{5-5} \cline{7-7} \cline{9-9}
\colhead{} & \colhead{} & \colhead{} & \colhead{} & \colhead{$log_{10}(M_*/M_{\odot})$} & \colhead{} & \colhead{} & \colhead{} & \colhead{} 
}

\startdata
                           & &                & &  8.5$-$9.0   & &  0.110 & &  $\ll$ 0.001 \\
                           & & Clusters  & &  9.0$-$9.5   & &  0.138 & &  $\ll$ 0.001  \\
  0.2 $< z <$ 0.5 & & vs.         & &  9.5$-$10.0  & &  0.156 & &  $\ll$ 0.001  \\
                           & & the Field  & & 10.0$-$10.5 & &  0.293 & &  $\ll$ 0.001  \\
                           & &                & & 10.5$-$11.0 & &  0.174 & &  $\ll$ 0.001  \\                              
  \cline{1-9}
                           & &                & &  8.5$-$9.0    & &  0.054 & &  $\ll$ 0.001  \\  
                           & &  Groups  & &  9.0$-$9.5    & &  0.133 & &  $\ll$ 0.001 \\
 0.2 $< z <$ 0.5  & &  vs.        & &  9.5$-$10.0   & &  0.243 & &  $\ll$ 0.001  \\
                           & & the Field     & & 10.0$-$10.5  & &  0.260 & &  $\ll$ 0.001  \\
                           & &                & & 10.5$-$11.0  & &  0.051 & &  0.078  \\                                                                           
 \cline{1-9}
                           & &  Clusters  & &  9.0$-$9.5   & &  0.126 & &  $\ll$ 0.001  \\ 
  0.5 $< z <$ 0.8 & &  vs.         & &  9.5$-$10.0  & &  0.195 & &  $\ll$ 0.001  \\
                           & & the Field       & & 10.0$-$10.5 & &  0.268 & &  $\ll$ 0.001  \\
                           & &                 & & 10.5$-$11.0 & &  0.143 & &  0.005  \\                                 
  \cline{1-9}
                           & &  Groups  & &  9.0$-$9.5   & &  0.063 & &  $\ll$ 0.001 \\    
 0.5 $< z <$ 0.8  & &  vs.        & &  9.5$-$10.0 & &  0.101 & &  $\ll$ 0.001 \\
                           & & the Field      & & 10.0$-$10.5 & &  0.195 & &  $\ll$ 0.001  \\
                           & &                & & 10.5$-$11.0 & &  0.096 & &  $\ll$ 0.001  \\                             
 \cline{1-9}
                           & &  Clusters & &  9.5$-$10.0   & &  0.273 & &  $\ll$ 0.001  \\   
  0.8 $< z <$ 1.1 & &  vs.         & & 10.0$-$10.5 & &  0.325 & &  $\ll$ 0.001  \\
                           & &  the Field      & & 10.5$-$11.0 & &  0.355 & &  $\ll$ 0.001  \\                                                       
  \cline{1-9}
                           & & Groups   & &  9.5$-$10.0   & &  0.115 & &  $\ll$ 0.001 \\   
 0.8 $< z <$ 1.1  & &  vs.         & & 10.0$-$10.5 & &  0.116 & &  $\ll$ 0.001  \\
                           & &  the Field      & & 10.5$-$11.0 & &  0.417 & &  $\ll$ 0.001  \\                                                                                  
 \enddata
 
\tablenotetext{a}{Output results based on $\href{https://docs.scipy.org/doc/scipy-0.14.0/reference/generated/scipy.stats.ks_2samp.html}{\textrm{stats.ks\_2samp (https://docs.scipy.org/doc/scipy-0.14.0/reference/generated/scipy.stats.ks\_2samp.html)}}$ function of python's Scipy package.}
\end{deluxetable*}

\section{Results and Discussion}
\subsection{sSFR Distribution}\label{sSFRDist}

To understand how galaxies populate in the sSFR space, we compute the distribution of the galaxy sSFR in different mass ranges, environments, and redshifts and plot the normalized sSFR against their total number in Figure~\ref{f2}. The purple and blue lines represent the sSFR distributions for the group and cluster galaxies, respectively, and the black dashed line denotes the field sample. The gold bar shows the green valley region, and the light blue line marks the position of the median sSFR of the star-forming galaxies in the field. We compute the mean values of the galaxy fraction with sSFR $<$ $-$10.0 in group, cluster, and field environments at different redshift ranges and list the results in Table~\ref{tab3}. From Table~\ref{tab3}, it is evident that the fraction exhibits a strong dependence on mass irrespective of  the environment or redshift.

From Figure~\ref{f2}, massive galaxies tend to be quiescent regardless of the environment, likely because massive galaxies may have been red and dead before being accreted to groups or clusters. In other words, the environmental quenching effect is not apparent for massive galaxies. By contrast, for low-mass galaxies, group or cluster galaxies exhibit a bimodal distribution, whereas field galaxies exhibit only one star-forming peak, implying that the quenching effect is more active in dense environments. At a fixed stellar mass, star-forming galaxies gradually turn into quiescent galaxies when evolving from high to low redshift as the peak of the sSFR distribution shifts from the star-forming to the quiescent regions. These findings are consistent with the conclusions in our previous work \citep{jian18} drawn through analysis of the fraction of quiescent galaxies.

The green valley region marked with the gold bar in Figure~\ref{f2} is defined previously in Section \ref{gv} on the SFR$-M_*$ plane. The definition suitably applies to the transition area. Notably, in the low-mass range, the gold bar appears at the valley between two peaks, indicating the star-forming and red sequence, respectively. The sample size and the fraction of galaxies in each category in Figure~\ref{f2} are summarized in Table~\ref{tab4}.

Based on the quenching timescale, the environmental quenching mechanisms can be broadly classified into two types: (1) fast quenching processes, such as ram pressure stripping \citep{gun72,dre83} and galaxy$-$galaxy interaction \citep{mih94}, where the star formation ceases over a short timescale of less than 1 Gyr, and (2) slow quenching effects [e.g., strangulation \citep{lar80,bal00}], where the removal of warm and hot gas can lead to a gradual reduction of cold gas supply and star formation quenching over a timescale of 1 Gyr or several gigayears.

 Star-forming galaxies that experience fast quenching quickly move to the inactive state without considerably changing the whole sSFR distribution of star-forming galaxies, leading to an sSFR distribution of star-forming populations similar to that of the field galaxies. On the other hand, in the case of slow quenching, the star-forming population is expected to result in a skewed distribution in sSFR toward low value and hence shows a systematically lower sSFR compared to that of field galaxies.  In other words, the properties of the distribution of the star-forming main sequence, such as their slope and amplitude on the SFR$-M_{*}$ plane, can be used as tools to infer the likely quenching mechanisms \citep[e.g.,][]{koy13,lin14}.

In our previous study \citep{lin14,jian17,jian18}, we measured the median sSFR of star-forming galaxies in different surroundings. We found an sSFR reduction of approximately 0.1\textendash 0.3 dex with mass dependence in the star-forming group or cluster galaxies compared with the star-forming field galaxies. The lower amplitude of the star-forming main sequence in groups/clusters thus suggests that the slow quenching plays at least partial (if not all) roles in dense environments.

In this study, we take a complementary approach to obtain a more comprehensive view of the sSFR distribution of star-forming galaxies by decomposing the full sSFR distribution into the star-forming and quiescent populations by fitting the full sSFR distribution with two Gaussian profiles. We note that our assumption preferentially ignores the non-gaussian sSFR pattern and thus limits the ability to identify asymmetries and tails in the star-forming population. Each Gaussian profile comprises three parameters (i.e., the peak amplitude, position, and width). We compute the error bars of the number count in each sSFR bins using the jack-knife resampling from eight subsamples. Based on the fitting function of Gnuplot,$\footnote{\url{http://www.gnuplot.info}}$ we obtain the best-fit result for six parameters and their asymptotic standard errors for the group, cluster, and field galaxies. We then normalize the star-forming Gaussian profiles with their peak values and plot the results in Figure~\ref{f3}. That is, the peak value of the star-forming sSFR profile is 1 in any environment.

To quantify the difference between the star-forming distributions, we perform two types of analyses. We first compute the best-fit peak value and its asymptotic standard error of the sSFR distribution in clusters (blue), groups (purple), and the field (black), as shown in the top left corner of each panel in Figure~\ref{f3}. The sSFR reduction and the significance of reduction between cluster and field galaxies (blue) and between group and field galaxies (purple) are listed on the middle left and bottom left of each subpanel in Figure~\ref{f3}, respectively, where the significance of the sSFR reduction is estimated by dividing the difference between two peak positions over the associated error. We find that the sSFR distribution of star-forming galaxies in the group or cluster environment shifts toward quiescence and exhibits a global sSFR reduction $\sim$ 0.1\textendash0.3 dex, consistent with results from previous works \citep{lin14,jian17,jian18}.

In the second analysis, we perform the two-sided Kolmogorov$-$Smirnov test (or K-S test) to evaluate the difference in the star-forming sSFR distributions between groups and field environments and between clusters and the field. The results are listed in Table~\ref{tab5}. We find that, in general, the $p-$value is small ($< 0.001)$ in any mass or redshift bins, suggesting that the difference between the two distributions is significant at $>$ 3.5$\sigma$.

\begin{figure*}
\includegraphics[scale=1.3]{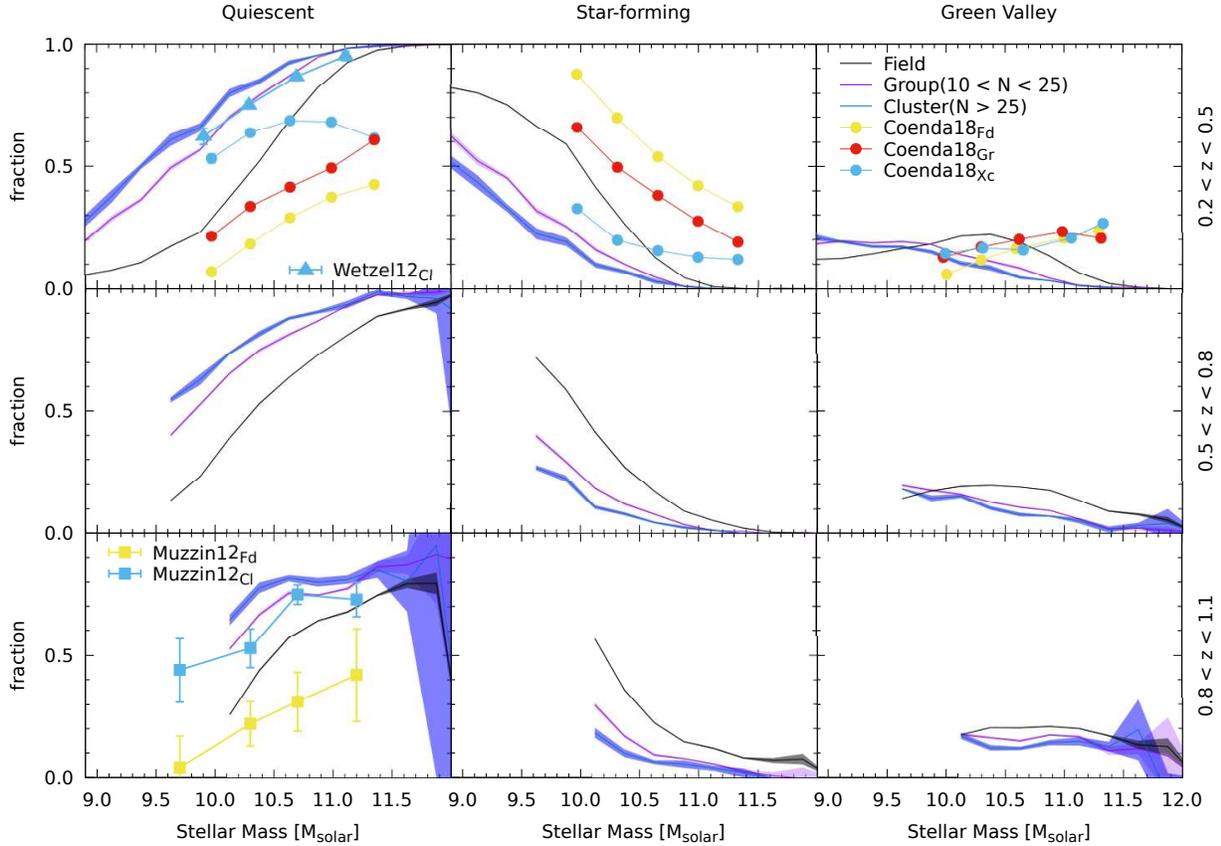}
\caption{Fraction of galaxies in three different environments (i.e., groups, clusters, and the field) are shown as functions of stellar mass and redshift. From left to right, quiescent, star-forming, and green valley galaxies are presented; from top to bottom, the redshift ranges are 0.2 $< z <$ 0.5, 0.5 $< z <$ 0.8, and 0.8 $< z <$ 1.1. At a fixed mass and environment, the sum of the fraction of galaxies in all three sequences equals 1. The shaded areas indicate 1$-\sigma$ regions from the jack-knife resampling by using eight subsamples. The solid blue, red, and yellow points represent the fraction of galaxies in an environment of X-ray cluster core (Xc), groups (Gr), and the field (Fd), respectively, from the results of \cite{coe18}.  The solid triangles mark the quiescent fraction for cluster galaxies from \cite{wet12}, and the blue and yellow solid squares represent the results from \cite{muz12} for cluster and field galaxies, respectively.}
\label{f4}
\end{figure*}

It is worth mentioning that many of the peaks of the star-forming sSFR are located in or below the green valley region. The reason for this result is that the star-forming main sequence flattens at the high-mass end $\sim$ $10^{10} \ M_{\odot}$ (see Figure~\ref{f1}), whereas our best-fit result of the SFR$-M_*$ relation for the star-forming galaxies is governed by low-mass galaxies, as described in Section~\ref{gv}. \cite{pan18} used the integral field spectroscopic observations from SDSS-IV MaNGA and demonstrated a flattening of the SFR-M$_*$ relation at the high-mass end. They found that the flattening is due to the growing regions in galaxies powered by nonstar formation sources generally with lower ionizing ability than the star formation sources. In other words, the flattening effect is likely due to a quenching effect, and the quenching also contributes to the increase in the number of green valley galaxies, although the quenching mechanism may not necessarily be associated with the environment.

The environmental impacts on sSFR among star-forming galaxies remains an subject under debate. Some studies have claimed an environmental ``independence'' of sSFR for star-forming galaxies \citep{pen10,mcg11,muz12,wij12,koy13}. On the other hand, an sSFR reduction of approximately 0.1\textendash 0.3 dex in the dense environments compared with the field environment has been identified in some other studies \citep{vul10,hai13,alb14,lin14,jian17,jian18}, with which our finding in this work is in good agreement. The difference is likely owing to the differences in the procedure of sample selection, the definition of SF galaxies, as well as the method used to measure the star formation rate \citep{koy18}.

\subsection{Fraction of Quiescent, Star-Forming, and Green Valley Galaxies}\label{fract}

Different quenching timescales also likely leave diverse imprints in the green valley fraction. In the case of fast quenching, star-forming galaxies quickly pass through the green valley region and become quiescent, without substantially increasing the green valley fraction. By contrast, if the quenching timescale is long (i.e., slow quenching), galaxies under the quenching may require more time to pass through the transition zone as a result of the increase in the fraction of green valley galaxies. Therefore, a comparison of green valley fractions in different environments  may provide insights into the timescale of the environmental quenching processes.

In Figure~\ref{f4}, the fraction of galaxies belonging to the quiescent (left), star-forming (middle), and green valley (right) populations are shown as a function of stellar mass in the low, medium, and high redshift bins from the top to bottom. Each panel displays the galaxy fraction in three different environments for comparison, namely clusters (blue), groups (purple), and the field (black). The quiescent and star-forming galaxies are those with the sSFR below and above the green valley region, respectively, slightly different from the result reported in our previous work \citep{jian18}, where no green valley galaxy was defined. At a fixed mass and environment, the sum of the fraction of galaxies in all the three populations equals one.

\begin{figure*}
\includegraphics[scale=1.3]{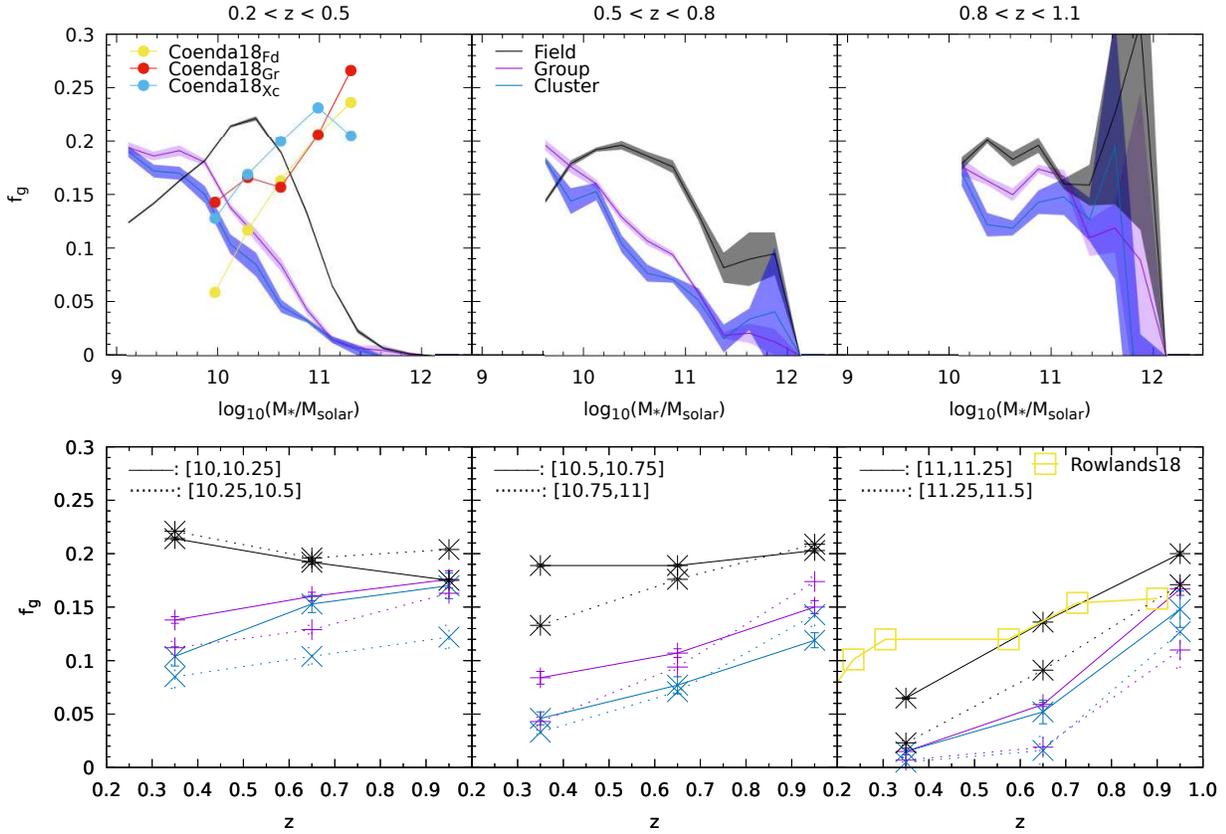}
\caption{Fraction of green valley galaxies ($f_g$) in the group, cluster, and field environments as a function of stellar mass (top) and redshift (bottom). The top panels show the zoomed-in plots of those in the right column (green valley) in Figure~\ref{f4}. The purple, green, and black lines denote the green valley galaxy fraction from the group, cluster, and field environments, respectively. The shaded region indicates 1$\sigma$ standard deviation from the jack-knife resampling. Evidently, the green valley fraction in this study exhibits a mass and redshift dependence, and its profile is distinct from that in \cite{coe18} at low-$z$.}
\label{f5}
\end{figure*}

For comparison,  we also present results from  \cite{wet12}, \cite{coe18}, and \cite{muz12} in Figure~\ref{f4}. \cite{coe18} constructed their galaxy samples from the Sloan Digital Sky Survey's seventh data release (DR7) with galaxy redshifts restricted to $z$ $\leq$ 0.15 and classified three sequences based on the NUV$-r$ color. The fraction of three populations in different environments from their results are plotted in the corresponding panel at low redshift, where blue, red, and yellow lines with solid points denote the result from X-ray cluster core (Xc), groups (Gr), and the field (Fd), respectively. Additionally, \cite{wet12} used the SDSS DR7 data to estimate the quiescent fraction in clusters by using sSFR as the classification indicator. Their results are denoted by the blue triangles for cluster mass between log mass 14.5 and 15.0 in the low-$z$ bin. Similarly, \cite{muz12} used sSFR as the indicator, and their quiescent fractions for cluster galaxies (blue squares) and field galaxies (yellow squares) in a redshift range of 0.85\textendash 1.2 are united in the high-$z$ bin.

For all environments, the quiescent fraction grows with an increase in mass, whereas the star-forming fraction exhibits an opposite trend. At low redshift, our result is consistent with the finding by \cite{wet12}. By contrast, the quiescent fractions in \cite{coe18} and \cite{muz12} are lower than those in this study in the same environment; a similar trend of an increase with an increase in mass is observed. Similarly, the star-forming fraction in the study reveals a lower value than that in \cite{coe18} for a fixed environment. In addition, quiescent fractions of group and cluster galaxies are similar but substantially higher than those of field galaxies, indicating an environmental dependence. Moreover, at a fixed mass, the quiescent fraction increases with a decreasing redshift, implying the Butcher\textendash Oemler effect \citep{but84}.

By contrast, the fraction of the green valley galaxies (or $f_g$) in general is weakly dependent on mass and redshift. The $f_g$s in groups or clusters are similar at any redshift and slightly decrease with an increase in mass. Moreover, the distribution of the green valley fraction in the field seems to peak at intermediate stellar mass between log mass 10 and 11, instead of exhibiting a monotonic dependence on stellar mass. However, because of the higher mass completeness limit at higher redshift, this feature is less certain at high redshifts. Additionally, the fraction of green valley galaxies in a dense environment intersects that in the field at log mass 10.0. Overall, $f_g$ is less than approximately 20$\%$ in all environments, consistent with the result reported by \cite{coe18} and \cite{row18}.

\begin{figure*}
\includegraphics[scale=1.3]{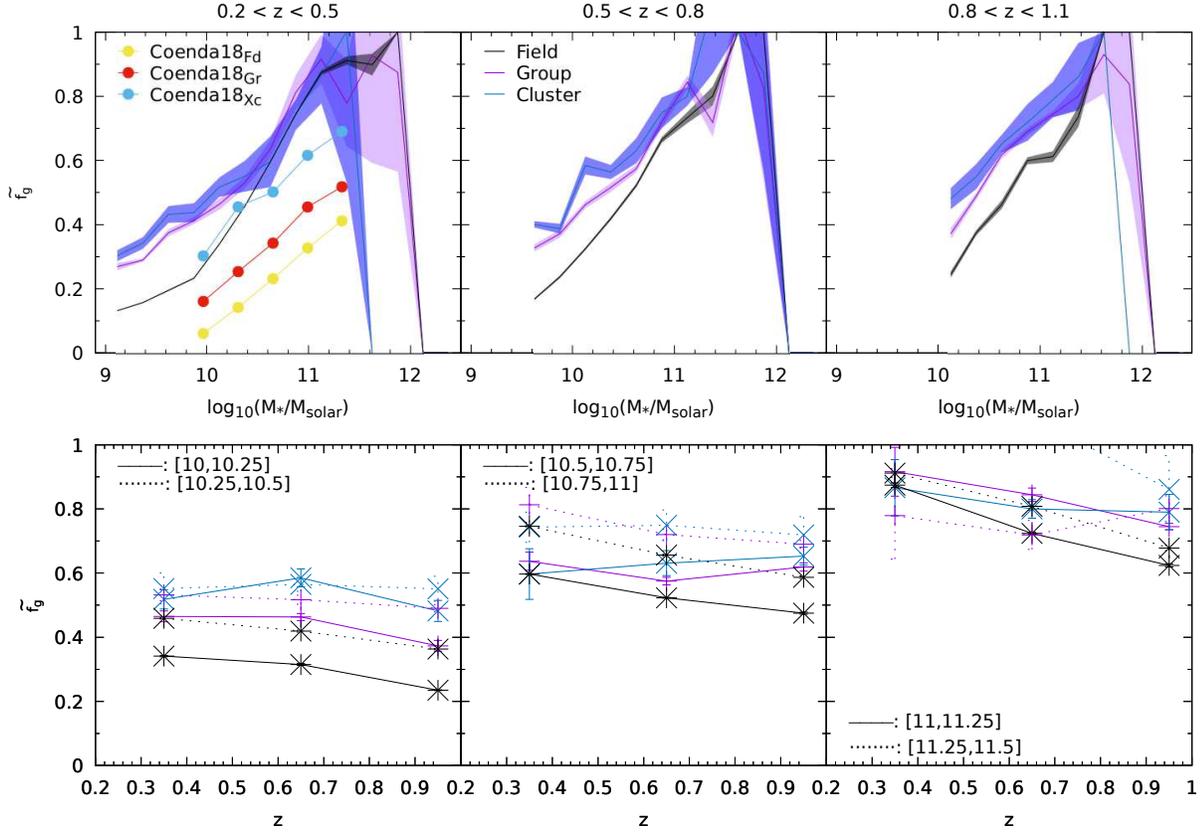}
\caption{Similar to Figure~\ref{f5}, we redefine the fraction of green valley galaxies by normalizing the number of green valley galaxies with the number of nonquiescent galaxies (or $\tilde{f_{g}}$) to reduce the effect from the dominance of quiescent galaxies. The $\tilde{f_g}$ in group (purple), cluster (blue), and field (black) environments are in comparisons. Overall, the $\tilde{f_{g}}$ depends strongly on mass and mildly on redshift and is greater for group or cluster galaxies than for field galaxies, suggesting an environmental quenching mechanism operating in the dense environments.}
\label{f6}
\end{figure*}

\subsection{Revisiting Green Valley Fraction}

We replot the relative fraction of green valley galaxies in Figure~\ref{f5} to further assess its dependence on mass (top) and redshift (bottom). The top panels indicate  that $f_g$ in groups and clusters is similar, but $f_g$ in the field displays a feature distinct from that in groups or clusters. In general, in the high-mass range in any environments $f_g$ decreases with decreasing redshift, which is likely due to the mass quenching leading to the insufficient supply of star-forming galaxies at low redshift. Compared with the results in \cite{coe18}, the $f_g$ at low-$z$ in the present study shows a different trend; $f_g$ in the present study decreases with mass, whereas that in their study increased with mass. The difference in the green valley fraction between field and group/cluster environments is at $>$ 2$\sigma$ confidence, for log galaxy mass $>$ 10.2 and $<$ 11.3 in any redshift range, where the significance of the difference is estimated by dividing the difference between two fractions over the associated error (i.e., the square root of the sum of the two fraction errors) in these two fractions.

 The bottom panels show a mild redshift evolution for low-mass field galaxies and a decrease in $f_g$ with a decreasing redshift for higher mass field galaxies. For group or cluster galaxies, $f_g$ also reveals a decreasing trend with the decreasing redshift, and the redshift dependence is slightly stronger for high-mass galaxies and weaker for low-mass galaxies. In addition, we plot the result from \cite{row18} with log$_{10}$($M/M_{\odot}$) $>$ 11.0 for comparison, and the consistency is evident between our and their results.

Notably, we found that $f_g$ in the field overcomes that in groups or clusters. This result is somewhat unexpected if galaxies are believed to experience additional quenching processes in a dense environment. A possible explanation is that star-forming galaxies can be believed to be the progenitors (or a reservoir) for producing green valley as well as quiescent galaxies. When the progenitors, especially for high-mass galaxies, are in shortage in groups or clusters, the relative number of green valley galaxies may be correspondingly smaller. Consequently, quiescent galaxies are quantitatively in dominance, whereas the green valley fraction in groups or clusters reveals a deficit compared with that in the field, as shown in the top panels in Figure~\ref{f5}.

To investigate the issue further, we redefine the fraction of green valley galaxies as the number of green valley galaxies over the number of nonquiescent galaxies (namely, the effective fraction of green valley galaxies, $\tilde{f_{g}}$) to remove the effect due to the dominance of the quiescent galaxies. The resultant $\tilde{f_{g}}$ as a function of mass (top) and redshift (bottom) is shown in Figure~\ref{f6}. The $\tilde{f_{g}}$ in all environments has a strong dependence on stellar mass. At all three redshift bins, the $\tilde{f_{g}}s$ in groups or clusters are comparable, and both are larger than that in the field for log mass approximately below 10.4 with a $>$ 3$\sigma$ confidence. That is, our result fits either the slow quenching scenario or the situation of fractionally more star-forming galaxies experiencing the fast quenching process. For both cases, the result suggests that an on-going environmental quenching effect likely has been acting in the dense environments since $z$ $\sim$ 1.

Combining the result from Section~\ref{sSFRDist}, we can thus infer that the environmental quenching involves some slow process(es) for galaxies with mass below log mass 11.0. The conclusion of the slow quenching effect in a dense environment is broadly consistent with the transition time of  approximately 1 Gyr reported by \cite{wong12}. The result is also compatible with results derived by \cite{phi19}, who demonstrated a green valley population dominated by galaxies with a timescale for star formation quenching of 2-4 Gyr. In addition, our result is in agreement with the finding by \citet{bel18} that green valley galaxies are a quasi-static population with a slow quenching process uniformly affecting the star formation rate over the entire galaxy.


\subsection{Systematic Effects}

In the present study, the radius for groups or clusters is fixed at 1.5 Mpc at all redshift ranges. To understand the aperture effect, we adjust the group and cluster boundary to 1.0 and 2.0 Mpc, respectively, and compare the fraction of three populations at the same redshift range for three apertures. We find that the fraction of green valley galaxies remains approximately unchanged at all redshift ranges. Furthermore, the fraction of quiescent galaxies in groups or clusters at the same redshift with a radius of 1.0 Mpc increases approximately by 5\% and with a radius of 2.0 Mpc decreases by only 5\%. Therefore, our results are not sensitive to the chosen aperture.

The Mostek method of deriving the SFR (i.e., Equation~\ref{eq2}) mentioned in Section~\ref{method} intrinsically introduces a scatter of approximately 0.19 dex for star-forming galaxies and 0.45 dex for quiescent galaxies \citep{mos12}. We perform Monte Carlo simulations to test the systematics caused by the uncertainty in the SFR of our sample. We mimic the sSFR distribution of galaxies by adopting two Gaussian profiles to represent the star-forming and quiescent populations, respectively. Given a value of the fraction for quiescent galaxies, the positions of the peak, and the widths, we can construct an unperturbed sSFR distribution for galaxies.

For every galaxy, we then randomly draw an excess value of sSFR based on the normal distribution with a width equal to the Mostek intrinsic scatter of 0.19 for star-forming or 0.45 for quiescent galaxies and add the excess to the galaxy sSFR. In this manner, we construct a perturbed sSFR distribution. We repeat the procedure 10,000 times to create an ensemble of the perturbed sSFR distribution and evaluate the bias and standard deviation of the median sSFR for two populations. We find that the bias is negligible ($\sim$ $<$ 0.04 dex), and therefore our comparison of the sSFR of star-forming galaxies between the field and groups/clusters is robust.

We also assess the effect of the intrinsic SFR scatters by using the Mostek method on the sSFR reduction of star-forming galaxies. We shift the whole star-forming population of approximately 0.2 dex toward quiescence to simulate the global sSFR reduction. In the meantime, we also randomly perturb the sSFR on individual star-forming galaxies with an uncertainty of 0.19 dex to mimic the intrinsic scatter in the SFR estimation by using the Mostek method. We find that for low-mass galaxies, the sSFR uncertainty does not change the initial reduction. However, for high-mass galaxies, the sSFR ambiguity smears the reduction signal and leads to little or no sSFR reduction. That is, the mass dependence of sSFR reduction (Figure~\ref{f3}) is likely due to the SFR uncertainty introduced by the Mostek method.  

We confirm that group or cluster galaxies experience a slow quenching process. However, the question remains whether galaxies in a dense environment quenched solely by the slow process can account for the difference in $f_q$ for group/cluster and field galaxies. To investigate this question, we recompute the quiescent fraction for field galaxies following the approach used in \cite{lin14} in the case that we globally reduce their sSFRs by an amount equal to the sSFR reduction $(\sim$ 0.1\textendash 0.3 dex) found in the star-forming main-sequence galaxies in groups/clusters compared with the field galaxies. Assuming that the change in quiescent fraction through the adjustment to account for the sSFR reduction in dense environments is purely due to the slow environmental quenching, we find that at low mass, only approximately 25\% of the excess quiescent fraction in groups (or clusters) relative to that in the field is contributed by the slow quenching process at all redshift ranges and environments. Even for high-mass galaxies, the contribution from slow quenching is approximately 50\%. That is, the environmental quenching effect in a dense environment cannot be attributed to only the slow quenching effect. 

 The excess quiescent fraction and the sSFR reduction in star-forming galaxies in groups or clusters relative to that in the field are commonly believed to be imprints left from the environmental quenching effect. However, we might also interpret the signatures as a consequence of the early formation of group or cluster galaxies. The green valley galaxies represent galaxies in the transition from star-forming to quiescence. The fraction of the green valley galaxies differ in different environments, indicating that the environmental quenching works in a dense surrounding. The scenario of the environmental quenching in groups or clusters is thus confirmed.

\section{Conclusions}
In this study, we use the DEmP photo-$z$ and CAMIRA catalogs from an HSC internal release in wide-field S17A to quantify the fraction of green valley galaxies in different environments and their redshift evolution. By determining the green valley line (i.e., the mean of the star-forming and red sequences on the SFR$-M_*$ plane), we define the green valley zone to be the region enclosed by 0.2 dex above and below the green valley line. The galaxies are then categorized into three types: star-forming, quiescent, and green valley. Three environments, namely groups, clusters, and the field in three redshift bins are constructed to discuss environmental and redshift dependence. The main results are summarized as follows:

\begin{enumerate}
\item From the sSFR distributions of galaxies in the group, cluster, and field environments, a global sSFR reduction in star-forming galaxies is observed in the dense environment relative to that in the field, consistent with the concept of a slow quenching process.   

\item The intrinsic fractions of the green valley galaxies in groups or clusters are comparable at any redshift and decrease with increasing mass and decreasing redshift. The green valley fractions in group/cluster environments differ from that in the field at low redshifts. Overall, the fraction of the green valley galaxies is less than approximately 20$\%$ in all environments.

\item By redefining the green valley fraction as the fraction of star-forming over nonquiescent galaxies, namely, the effective green valley fraction ($\tilde{f_{g}}$), we find that $\tilde{f_{g}}$s in cluster, group, and field galaxies increase rapidly with stellar mass but exhibit a mild redshift dependence. In addition, $\tilde{f_{g}}$ in groups or clusters is higher than that in the field at any mass, implying that an environmental quenching process operates in the dense environment. Combining the result from Figure~\ref{f3}, we conclude that an ongoing slow quenching mechanism is acting in the dense environment since $z$ $\sim$ 1.

\end{enumerate}

In this paper, we discussed the environmental impact on the star formation quenching under various mass and redshift conditions. However, the environmental dependence of star formation quenching is not only on the cluster halo mass but also on other factors, such as cluster-centric radius and galaxy local density. The relative importance of these parameters can be more efficiently quantified by probing the star formation quenching status in cluster galaxies in terms of these parameters concurrently. The final HSC-wide sample will be five times larger than the sample size used in this work, allowing the breaking of degeneracies between different environmental processes.

\emph{Acknowledgments}-
This work was supported by the Academia Sinica under
the Career Development Award CDA-107-M03 and the Ministry of Science \& Technology of Taiwan
under the grants MOST 107-2119-M-001-024 and MOST 108-2628-M-001-001-MY3.
M. Oguri was supported in part by World Premier International Research Center Initiative (WPI Initiative), MEXT, Japan, and JSPS KAKENHI grant No. JP15H05892 and JP18K03693. The Hyper Suprime-Cam (HSC) collaboration includes the astronomical communities of Japan and Taiwan and Princeton University.  The HSC instrumentation and software were developed by the National Astronomical Observatory of Japan (NAOJ), the Kavli Institute for the Physics and Mathematics of the Universe (Kavli IPMU), the University of Tokyo, the High Energy Accelerator Research Organization (KEK), the Academia Sinica Institute for Astronomy and Astrophysics in Taiwan (ASIAA), and Princeton University.  Funding was contributed by the FIRST program from Japanese Cabinet Office, the Ministry of Education, Culture, Sports, Science and Technology (MEXT), the Japan Society for the Promotion of Science (JSPS),  Japan Science and Technology Agency  (JST),  the Toray Science  Foundation, NAOJ, Kavli IPMU, KEK, ASIAA,  and Princeton University.

\end{document}